\documentclass[pra,twocolumn,amsmath,amssymb,amsfonts,showkeys,showpacs,preprintnumbers,floatfix]{revtex4-1}
\usepackage{bm,graphicx,xcolor,hyperref,multirow,rotating}

\newcommand{\app}{\approx}
\newcommand{\eps}{\varepsilon}
\newcommand{\om}{\omega}
\newcommand{\Ec}{E_{\rm c}}

\begin{document}
\bibliographystyle{apsrev}

\title{Correlation energy of two electrons in the high-density limit}

\author{Pierre-Fran\c{c}ois Loos}
\author{Peter M. W. Gill}
\email{peter.gill@anu.edu.au}
\affiliation{Research School of Chemistry, Australian National University, Canberra, ACT 0200, Australia}
\date{\today}

\begin{abstract}
We consider the high-density-limit correlation energy $\Ec$ in $D \ge 2$ dimensions for the $^1S$ ground states of three two-electron systems: helium (in which the electrons move in a Coulombic field), spherium (in which they move on the surface of a sphere), and hookium (in which they move in a quadratic potential).  We find that the $\Ec$ values are strikingly similar, depending strongly on $D$ but only weakly on the external potential.  We conjecture that, for large $D$, the limiting correlation energy $\Ec \sim -\delta^2/8$ in any confining external potential, where $\delta = 1/(D-1)$.
\end{abstract}


\maketitle

\section{Introduction}
The concept of electron correlation energy ($\Ec$) is an old and important one, first introduced by Wigner \cite{Wigner34} and later defined by L\"owdin \cite{Lowdin59} as the error
\begin{equation}
	\Ec = E - E_{\rm HF}
\end{equation}
of the Hartree-Fock (HF) model.  Understanding and calculating $\Ec$ is one of the most important and difficult problems in quantum chemistry and molecular physics.

The observation that HF theory is useful for the prediction of molecular structure \cite{HRSPbook} suggests that $\Ec$ often depends only weakly on the external potential.  To explore this, we have studied the correlation energy $\Ec(D, m, Z)$ of two opposite-spin electrons, confined to a $D$-dimensional space and moving in an external potential $Z^{m+2} V(r)$ where $V(r) \propto r^m$.

We consider three such systems.  In $D$-helium, the electrons move in the Coulomb potential $V(r) = -1/r$.  In $D$-spherium \cite{Ezra82}, they move in a constant potential $V(r) = r^0$ on a $D$-sphere of radius $1/Z$.  (In 2-spherium, for example, this is the surface of a three-dimensional ball.)  In $D$-hookium (also known as Hooke's atom or harmonium) \cite{Kestner62}, they move in the harmonic potential $V(r) = r^2/2$.

After the length scaling $r \leftarrow Zr$, the Hamiltonians of the three systems reduce to the form \cite{Louck60, Herschbach86}
\begin{equation}
	\hat{H} = - \frac{\nabla_1^2}{2} - \frac{\nabla_2^2}{2} + V(r_1) + V(r_2) + \frac{1}{Z r_{12}}
\end{equation}
where $r_{12} = |\bm{r}_1-\bm{r}_2|$.  Following Hylleraas \cite{Hylleraas30}, perturbation theory can then be used to expand both the exact and HF energies as series in $1/Z$, yielding
\begin{align}
	E		& = E_0(D,m) Z^2	+ E_1(D,m) Z + E_2(D,m) 		+ \mathcal{O}(Z^{-1})	\label{Eex}	\\
	E_{\rm HF}	& = E_0(D,m) Z^2	+ E_1(D,m) Z + E_2^{\rm HF}(D,m)	+ \mathcal{O}(Z^{-1})	\label{EHF}
\end{align}
Many workers have investigated the energies
\begin{equation} \label{EcDm}
	\Ec(D,m) = \lim_{Z\to\infty} \Ec = E_2(D,m) - E_2^{\rm HF}(D,m)
\end{equation}
that arise in the high-density limit.  Studies of the helium-like ions \cite{Hylleraas30, Baker90}, for example, showed that $\Ec(3,-1) \app 47$ mh (millihartrees) and several groups have noted that the limits for 3-spherium [$\Ec(3,0) \app 48$ mh] \cite{Quasi09} and 3-hookium [$\Ec(3,2) \app 50$ mh] \cite{White70, Katriel05, HookCorr05} are similar.

Two-dimensional systems have also been studied and, although the limiting energies for 2-helium [$\Ec(2,-1) > 212$ mh] \cite{Loeser87}, 2-spherium [$\Ec(2,0) \app 227$ mh] \cite{Seidl07, TEOAS1} and 2-hookium [$\Ec(2,2) > 162$ mh] \cite{Rod01} are several times greater than their $D=3$ analogs, they appear similar to one another.  Conversely, those for 4-helium [$\Ec(4,-1) > 18$ mh] and higher heliums are much smaller \cite{Loeser87}.

Such results suggest that the limiting correlation energies are similar, not only for $D=3$ as previously reported, but also for other $D$.  It leads to the idea that \emph{the correlation energy of two electrons in the high-density limit depends strongly on the dimensionality of the space in which they move, but weakly on the external potential}.  To explore this, we have calculated the limiting correlation energies of the $^1S$ ground states of helium, spherium and hookium for $D=2,3,\ldots,8$.  We use atomic units throughout.

\section{Helium}
The one-electron Hamiltonian in $D$-helium is
\begin{equation}
	\Hat{H}_0 = - \frac{1}{2} \left[ \frac{d^2}{dr^2} + \frac{D-1}{r} \frac{d}{dr} \right] - \frac{1}{r}
\end{equation}
and the zeroth-order wave function is
\begin{equation}
	\Psi_0(\bm{r}_1,\bm{r}_2) = \frac{4^D}{(D-1)^{D}\Gamma(D)} \exp \left( - \frac{2r_1+2r_2}{D-1}\right)
\end{equation}
The $E_0$ and $E_1$ values are
\begin{align}
	E_0(D,-1)	& = - \frac{4}{(D-1)^2}	\\
	E_1(D,-1)	& = \frac{4}{(D-1)^2} \frac{\Gamma\left(D+\frac{1}{2}\right) \Gamma\left(\frac{D+1}{2}\right)}
														{\Gamma(D+1) \Gamma \left(\frac{D}{2}\right)}
\end{align}
where $\Gamma$ is the Gamma function \cite{ASbook}.  

$E_2$ values were computed using the Hylleraas method \cite{Hylleraas30}.  We adopted the length and energy scaling of Herrick and Stillinger \cite{Stillinger75} and used the Hylleraas basis functions \cite{Hylleraas30}
\begin{equation}
	\psi_{n,l,m} = s^n t^l u^m \exp(-s/2)
\end{equation}
where $s = r_1 + r_2 $, $t = r_1 - r_2$ and $u = r_{12}$. The second-order energy, which minimizes the Hylleraas functional, is then given by
\begin{equation}
	E_2(D,-1) = - \frac{1}{2} \mathbf{b^{\rm T}A^{-1}b}													 \label{E2-he}
\end{equation}
where
\begin{gather}
	\mathbf{A}_{\om_1 \om_2} = \mathbf{M}_{\om_1 \om_2} - \frac{D-1}{2} \mathbf{L}_{\om_1 \om_2}
																	- 2 E_0 \mathbf{S}_{\om_1 \om_2}	\label{Hy-A}	\\
	\mathbf{b}_{\om} = 2 E_1 \mathbf{S}_{0 \om} - \frac{D-1}{2} \mathbf{V}_{0 \om}						\label{Hy-B}
\end{gather}
with $\om=(n,2l,m)$.  In \eqref{Hy-A} and \eqref{Hy-B}, $\mathbf{M}$, $\mathbf{L}$, $\mathbf{S}$ and $\mathbf{V}$ are the kinetic, electron-nucleus, overlap and repulsion matrices, respectively.  Details can be found elsewhere \cite{Hylleraas64, Stillinger75}.  The Hylleraas basis was progressively enlarged by increasing the maximum values of $n$, $l$ and $m$ until the most difficult case ($D=2$) converged to 6 digits.

Although the $E_2$ value for 3-helium has been studied in great detail (as in, for example, the work of Morgan and co-workers \cite{Baker90, Kutzelnigg92}), the only other helium whose $E_2$ value has been reported \cite{Stillinger75} (by exploiting interdimensional degeneracies \cite{Herrick75, Goodson91}) is 5-helium.

Although Loeser and Herschbach have investigated the dimensional dependence of the HF energy of helium \cite{Loeser86, Herschbach86}, $E_2^{\rm HF}(D,-1)$ has been reported \cite{Linderberg61} only for $D=3$.  All values can be found using the generalization
\begin{gather}
	E_2^{\rm HF}(D,-1) = - \int_0^\infty \frac{W(r)^2}{r^{D-1} \, \Psi_0(r)^2}\,dr	\label{E2HF-he}	\\
	W(r) = 2 \int_0^r [J(x) - E_1] \, \Psi_0(x)^2 \, x^{D-1} \, dx
\end{gather}
of the Byers-Brown--Hirschfelder equations \cite{ByersBrown63}, where
\begin{equation}
	J(r) = \int_0^\infty \frac{\Psi_0(r)^2}{\max(r,x)} F\left[ \frac{3-D}{2},\frac{1}{2},\frac{D}{2},\alpha^2 \right] x^{D-1} dx
\end{equation}
$\alpha = \frac{\min(x,r)}{\max(x,r)}$ and $F$ is the hypergeometric function \cite{ASbook}.  For odd $D$, this yields simple expressions such as
\begin{align}
	E_2^{\rm HF}(3,-1)	& = + \frac{9}{32} \ln \frac{3}{4} - \frac{13}{432}	\\
	E_2^{\rm HF}(5,-1)	& = - \frac{903}{1024} \ln \frac{3}{4} - \frac{35\,213}{124\,416}
\end{align}

Eqs~\eqref{E2-he} and \eqref{E2HF-he} yield the large-$D$ expansions \cite{Mlodinow81}
\begin{align}
	E_2(D,-1)			& \sim - \frac{5}{8}\delta^2 - \frac{31}{384}\delta^3	+ \ldots		\\
	E_2^{\rm HF}(D,-1)	& \sim - \frac{1}{2}\delta^2 + \frac{3}{32}\delta^3	+ \ldots		\\
	\Ec(D,-1)			& \sim - \frac{1}{8}\delta^2 - \frac{67}{384}\delta^3	+ \ldots		\label{Asy-he}
\end{align}
where, following previous work \cite{Yaffe83,Doren85}, we use $\delta = 1/(D-1)$.

\section{Spherium}
The zeroth-order Hamiltonian of $D$-spherium is
\begin{equation}
	\Hat{H}_0 =  - \frac{d^2}{d\theta^2} - (D-1) \cot \theta \frac{d}{d\theta}
\end{equation}
(where $\theta$ is the inter-electronic angle) and the associated eigenfunctions and eigenvalues are, respectively,
\begin{gather}
	\Psi_n(\theta) = \mathcal{N}\,C_n^{\frac{D-1}{2}}(\cos\theta)	\\
	\eps_n = n (n+D-1)
\end{gather}
where $C_n^{\frac{D-1}{2}}$ is a Gegenbauer polynomial \cite{ASbook} and
\begin{equation}
	\mathcal{N} = \sqrt{\frac{2^{D-3} (2n+D-1) \Gamma \left(\frac{D-1}{2}\right)^2 \Gamma (n+1)}{\pi \Gamma (n+D-1)}}
\end{equation}
Using the partial-wave expansion of $r_{12}^{-1}$, one finds
\begin{equation}
	\left< C_0^{\frac{D-1}{2}} \Big| r_{12}^{-1} \Big| C_n^{\frac{D-1}{2}} \right> = \frac{(n+1)_{D-2}}{(n+\frac{1}{2})_{D-1}}
\end{equation}
where $(a)_n$ is a Pochhammer symbol \cite{ASbook} and, therefore,
\begin{equation}
	E_1(D,0) = \frac{\Gamma(D-1) \Gamma\left(\frac{D+1}{2}\right)}{\Gamma\left(D-\frac{1}{2}\right) \Gamma\left(\frac{D}{2}\right)}
\end{equation}
The second-order energy is given by
\begin{align} \label{E2-sp}
	E_2(D,0)	& = \sum_{n=1}^{\infty} \frac{\left< \Psi_0 \left| r_{12}^{-1} \right| \Psi_{n}\right>^2}{\eps_0 - \eps_{n}}		\notag	\\
			& = - \frac{\Gamma(D)}{4\pi} \frac{\Gamma \left(\frac{D-1}{2}\right)^2}{\Gamma \left(\frac{D}{2}\right)^2}		
			\sum_{n=1}^\infty \frac{(n+1)_{D-2}}{(n+\frac{1}{2})_{D-1}^2} \left[\frac{1}{n}+\frac{1}{n+D-1}\right]
\end{align}
which reduces to generalized hypergeometric functions.  It is easy to show \cite{TEOAS1} that $E_0(D,0) = 0$ and
\begin{equation} \label{E2HF-sp}
	E_2^{\rm HF}(D,0) = 0
\end{equation}

The $E_2$ (and thus $\Ec$) value for 2-spherium was recently reported by Seidl \cite{Seidl07}.  However, simple expressions for any $D$ can be obtained from Eq.~\eqref{E2-sp}.  For example,
\begin{subequations}
\begin{align}
	\Ec(2,0)		& = 4 \ln 2 - 3							\\
	\Ec(3,0)		& = \frac{4}{3} - \frac{368}{27 \pi^2}		\\
	\Ec(4,0)		& = \frac{64}{75} \ln 2 - \frac{229}{375}	\\
	\Ec(5,0)		& = \frac{24}{35} - \frac{2\,650\,112}{385\,875 \pi^2}
\end{align}
\end{subequations}
Eq.~\eqref{E2-sp} also yields the large-$D$ expansion
\begin{equation} \label{Asy-sp}
	\Ec(D,0) \sim - \frac{1}{8}\delta^2 - \frac{21}{128}\delta^3 + \frac{21}{512}\delta^4 + \ldots
\end{equation} 

\section{Hookium}
The one-electron Hamiltonian in $D$-hookium is
\begin{equation}
	\Hat{H}_0 = - \frac{1}{2} \left[ \frac{d^2}{dr^2} + \frac{D-1}{r} \frac{d}{dr} \right] + \frac{r^2}{2}
\end{equation}
and the zeroth-order wave functions are
\begin{equation}
	\Psi_\ell(\bm{r}_1,\bm{r}_2) = \prod_{k=1}^{D} \psi_{a_k} (x_{1,k}) \psi_{b_k} (x_{2,k})
\end{equation}
where $x_{i,k}$ is the $k$th cartesian coordinate of electron $i$, and $a_k$ and $b_k$ are non-negative integers.  The orbitals are the one-dimensional harmonic oscillator wave functions
\begin{equation}
	\psi_a(x) = \sqrt{2^a a! \pi^{1/2}} H_a(x) \exp(-x^2/2)
\end{equation}
where $H_a$ is the $a$th Hermite polynomial \cite{ASbook}.   The energy difference between the eigenstates are given by
\begin{equation}
	\eps_\ell - \eps_0 = \sum_{k=1}^D (a_k+b_k) = 2n	\label{excitation}
\end{equation}
where $2n$ is the excitation level, \textit{i.e.}~the number of nodes in $\Psi_\ell$.  It is not difficult to show that $E_0(D,2) = D$ and
\begin{equation}
	E_1(D,2) = \frac{1}{\sqrt{2}} \frac{\Gamma(\frac{D-1}{2})}{\Gamma(\frac{D}{2})}
\end{equation}

Both $E_2$ and $E_2^{\rm HF}$ can be found by direct summation \cite{HookCorr05}, as in Eq.~\eqref{E2-sp}.  The sum includes all single and double excitations for $E_2$, but only singles for $E_2^{\rm HF}$.  The integral $\left< \Psi_0 \left| r_{12}^{-1} \right| \Psi_\ell \right>$ vanishes unless all of the $a_k+b_k$ are even and, in that case, it is given by
\begin{equation} \label{int-ho}
	\left< \Psi_0 \left| r_{12}^{-1} \right| \Psi_\ell \right> =
	\frac{1}{\sqrt{2 \pi}} \frac{\Gamma \left(\frac{D-1}{2}\right) \Gamma \left(n+\frac{1}{2}\right) }{\Gamma(n+2)} 
	\prod_{k=1}^D \frac{i^{a_k-b_k} }{\sqrt{\pi a_k! b_k!}} \Gamma \left(\frac{a_k+b_k+1}{2}\right)
\end{equation}
In this way, one eventually finds
\begin{gather}
	E_2(D,2)			= -\frac{\Gamma \left(\frac{D-1}{2}\right)^2}{4\,\Gamma\left(\frac{D}{2}\right)^2}
 						\sum_{n=1}^\infty \frac{\left(\frac{1}{2}\right)_n^2}{\left(\frac{D}{2}\right)_n} \frac{1}{n!\,n}		\label{E2-ho}	\\
	E_2^{\rm HF}(D,2)	= -\frac{\Gamma \left(\frac{D-1}{2}\right)^2}{2\,\Gamma\left(\frac{D}{2}\right)^2}
	 					\sum_{n=1}^\infty \frac{\left(\frac{1}{2}\right)_n^2}{\left(\frac{D}{2}\right)_n} \frac{(1/4)^n}{n!\,n}	\label{E2HF-ho}
\end{gather}
which reduce to generalized hypergeometric functions.

$E_2(3,2)$ has been derived by several groups \cite{White70, Cioslowski00, HookCorr05} but the energies for other $D$ have not been reported before.  All can be found in closed form and the first few are
\begin{subequations}
\begin{align}
	E_2(2,2)	& = 2G - \pi \ln 2										\\
	E_2(3,2)	& = 1 - \frac{2}{\pi}(1+\ln 2)								\\
	E_2(4,2)	& = \frac{1}{4} \left[ 2G - \pi \ln 2 + 1 - \frac{\pi}{4} \right]	\\
	E_2(5,2)	& = \frac{5}{9} - \frac{8}{27\pi} (4+3 \ln 2)
\end{align}
\end{subequations}
where $G$ is Catalan's constant \cite{ASbook}.  Similar remarks pertain to the HF energies with odd $D$, such as
\begin{subequations}
\begin{align}
	E_2^{\rm HF}(3,2)	& = \frac{4}{3} - \frac{4}{\pi} \left[ 1 + \ln(8-4\sqrt{3}) \right]	\\
	E_2^{\rm HF}(5,2)	& = \frac{8}{27} - \frac{8}{27\pi} \left[ 8 - 3\sqrt{3} + 6 \ln(8-4\sqrt{3}) \right]
\end{align}
\end{subequations}
Eqs \eqref{E2-ho} and \eqref{E2HF-ho} also yield the large-$D$ expansions
\begin{align}
	E_2(D,2)			& \sim - \frac{1}{4}\delta^2 - \frac{5}{32}\delta^3		+ \frac{3}{64}\delta^4		+ \ldots		\\
	E_2^{\rm HF}(D,2)	& \sim - \frac{1}{8}\delta^2 + \frac{7}{256}\delta^3	+ \frac{21}{1024}\delta^4	+ \ldots		\\
	\Ec(D,2)			& \sim - \frac{1}{8}\delta^2 - \frac{47}{256}\delta^3	+ \frac{27}{1024}\delta^4	+ \ldots	\label{Asy-ho}
\end{align}

\begin{table*}
\caption{\label{tab:E2} Second-order energies and limiting correlation energies in two-electron systems.}
\begin{ruledtabular}
\begin{tabular}{lcccccccc}
	System		&	$m$	&	$D=2$		&	$D=3$		&	$D=4$		&	$D=5$		&	$D=6$		&	$D=7$		&	$D=8$		\\
	\hline
				&			&			\multicolumn{7}{c}{Second-order exact energies, $-E_2(D,m)$, from \eqref{Eex}}					\\
	Helium		&	$-1$	&	0.632740	&	0.157666 	&	0.070044	&	0.039395 	&	0.025208	&	0.017501	&	0.012854	\\
	Spherium	&	$0$		&	0.227411 	&	0.047637 	&	0.019181	&	0.010139	&	0.006220	&	0.004189	&	0.003007	\\
	Hookium	&	$+2$	&	0.345655	&	0.077891 	&	0.032763	&	0.017821	&	0.011153	&	0.007622	&	0.005533	\\
	\hline
				&			&			\multicolumn{7}{c}{Second-order HF energies, $-E_2^{\rm HF}(D,m)$, from \eqref{EHF}}			\\
	Helium		&	$-1$	&	0.412607	&	0.111003 	&	0.051111	&	0.029338	&	0.019020	&	0.013325	&	0.009852	\\
	Spherium	&	$0$		&	0 			&	0 			&	0			&	0			&	0			&	0			&	0			\\
	Hookium	&	$+2$	&	0.106014	&	0.028188 	&	0.012904	&	0.007382	&	0.004776	&	0.003342	&	0.002469	\\
	\hline
				&			&			\multicolumn{7}{c}{Limiting correlation energies $-\Ec(D,m)$, from \eqref{EcDm}}					\\
	Helium		&	$-1$	&	0.220133	&	0.046663	&	0.018933	&	0.010057	&	0.006188	&	0.004176	&	0.003002	\\
	Spherium	&	$0$		&	0.227411	&	0.047637	&	0.019181	&	0.010139	&	0.006220	&	0.004189	&	0.003007	\\
	Hookium	&	$+2$	&	0.239641	&	0.049703	&	0.019860	&	0.010439	&	0.006376	&	0.004280	&	0.003065	\\
\end{tabular}
\end{ruledtabular}
\end{table*}

\section{Results and Discussion}
Numerical values of $E_2$, $E_2^{\rm HF}$ and $E_c$, for $D=2,\ldots,8$ and $m=-1$, 0 and 2 are reported in Table \ref{tab:E2}.  The $E_2$ values for helium were found by the Hylleraas technique described in Section II.  Other results were obtained from Eqs \eqref{E2HF-he}, \eqref{E2-sp}, \eqref{E2HF-sp}, \eqref{E2-ho} and \eqref{E2HF-ho}.

As $m$ increases (for constant $D$), although the exact and HF energies decrease in magnitude from helium to spherium and then increase from spherium to hookium, the correlation energies always increase.  However, the smallness of that increase is striking;  $\Ec$ is almost independent of $m$, especially for large $D$.  The correlation energies of helium and hookium differ by only 8\% for $D=2$, and this drops to just 2\% for $D = 8$.

As $D$ increases (for constant $m$), all of the energies decrease dramatically and the correlation energies fall by almost two orders of magnitude between $D=2$ and $D=8$.  Herrick and Stillinger have explained this in $D$-helium \cite{Stillinger75} by observing that the Jacobian $(r_1 r_2)^{D-1}$ creates a ``dimensionality barrier'' that keeps both electrons far from the nucleus and therefore allows them to avoid each other more easily when $D$ is large.  Similar arguments apply to $D$-spherium and $D$-hookium and, presumably, in general.

The observed dependence of the correlation energy on $D$ and $m$ is consistent with the large-$D$ expansions \eqref{Asy-he}, \eqref{Asy-sp} and \eqref{Asy-ho}, all of which take the form
\begin{equation} \label{Asy}
	\Ec(D,m) \sim - \delta^2 / 8 - C\,\delta^3
\end{equation}
where the coefficient $C \app 1/6$ varies slowly with $m$.  Such an expression implies that $\Ec$ depends primarily on the dimensionality of space in which the electrons move but with a small correction from the shape of the confining external potential.

We conjecture that Eq.~\eqref{Asy} is true for all confining external potentials $V(r)$.  To explore this, it would be useful to extend Table \ref{tab:E2} to include ``airium'' [$V(r) = r$], ``ballium'' \cite{Alavi05} [$V(r)=r^\infty$] and other such systems. These studies will be reported elsewhere.

\begin{acknowledgments}
We thank Andrew Gilbert for several stimulating discussions and Yves Bernard for helpful comments on this manuscript.  P.M.W.G. thanks the APAC Merit Allocation Scheme for a grant of supercomputer time and the Australian Research Council (Grants DP0664466 and DP0771978) for funding.
\end{acknowledgments}

\end{document}